\documentclass {JHEP3}

\def\IR{\relax{\rm I\kern-.18em R}}

\def\CN{{\cal N}}

\def\IR{\relax{\rm I\kern-.18em R}}

\def\CN{{\cal N}}

\author{Aybike \c{C}atal-\"{O}zer \\
{\it School of Mathematics, Trinity College Dublin\\
Dublin 2 IRELAND \\ \underline{e-mail}: aybike@maths.tcd.ie} }

\abstract{We examine the solution generating symmetries by which
Lunin and Maldacena have generated  the gravity duals of
$\beta$-deformations of certain field theories. We identify the
$O(2,2,\IR)$ matrix, which acts on the background matrix $E=g+B$,
where $g$ and $B$ are the metric and the B-field of the undeformed
background, respectively. This simplifies the calculations and
makes some features of the deformed backgrounds more transparent.
We also find a new three-parameter deformation of the
Sasaki-Einstein manifolds $T^{1,1}$ and $Y^{p,q}$. Following the
recent literature on the three-parameter deformation of $AdS_5
\times S^5$, one would expect that our new solutions should
correspond to non-supersymmetric marginal deformations of the
relevant dual field theories.}

\keywords{String duality, Gauge-gravity correspondence}

\title{Lunin-Maldacena Deformations with Three Parameters}

\begin{document}


\section{Introduction}

In a recent paper, Lunin and Maldacena studied a specific type of
deformation of certain field theories and presented a method to
find the gravity duals of the deformations \cite{LM}. The type of
deformations they consider apply to both conformal and
non-conformal field theories and is such that the resulting theory
preserves a global $U(1) \times U(1)$ symmetry. The effect of the
deformation in the Lagrangian is to introduce phases depending on
the order of fields charged under this global symmetry. Such
deformations are sometimes called $\beta$ deformations, as they
are characterized by a complex parameter, usually called $\beta$.
In the particular case when $\beta$ is real, they are usually
called $\gamma$ deformations. In this paper, we will be interested
only in real deformations. The prescription  Lunin and Maldacena
gave works when the global $U(1) \times U(1)$ is realized
geometrically on the gravity side. In this case, the geometry of
the gravity background contains a 2-torus and the new gravity
solutions corresponding to the deformed field theory are generated
by using an $SL(2,\IR)$ symmetry associated with the 2-torus.
However, this $SL(2,\IR)$ is not to be confused with the geometric
$SL(2,\IR)$ acting on the complex structure modulus. It rather
acts on the K\"{a}hler modulus $\rho$, whose real and imaginary
parts are the value of the $B$ field on the 2-torus and the volume
of the 2-torus, respectively. This is the non-geometric part of
the symmetry group $O(2,2,\IR)$ associated with the two commuting
isometries. We will discuss this further in Section 2.

Later in \cite{F}, Frolov examined the deformed gravity solutions
of \cite{LM} and showed that the classical bosonic string theory
in this background exhibits a Lax pair and as such, is integrable.
The Bethe ansatz for the integrable spin chain was presented in
\cite{FRT1} and the string Bethe equations were solved explicitly.
Again in \cite{F}, Frolov generalized the deformed gravity
solutions of \cite{LM} by introducing two more parameters and
found a 3-parameter family of deformations of $AdS_5 \times S^5$,
which he proposed as the gravity dual of a non-supersymmetric
marginal deformation of $\CN=4$ SYM, at least in the large N
limit. The integrable spin chain corresponding to this background
and the Bethe ansatz were constructed in \cite{beisert}. Further
checks of AdS/CFT correspondence for the new 3-parameter family of
deformations were performed in \cite{FRT2}. For more on
$\beta$-deformations and their gravity duals, see for example
\cite{LS,Aharony:1999ti,cherkis,gursoy,ahn,gauntlett3,deMelloKoch:2005vq,mateos,gursoy2,Penati:2005hp,kumar}.
Also see the very recent \cite{frolovnew} for more on
Lunin-Maldacena deformations with several parameters.

One of our aims in this paper is to have a closer look at the
underlying symmetries by which the new gravity solutions have been
generated in \cite{LM}. In particular, we will find the
$O(2,2,\IR)$ matrix acting on the background matrix $E = g+B$, so
that we obtain the solutions in an easier way. This also makes
some features of the new backgrounds more transparent. For
example, the transformation of the coordinates and the momenta and
winding charges, which Frolov utilized in writing the Lax pair for
the deformed $AdS_5 \times S^5$, follows easily from the
properties of the $O(2,2,\IR)$ matrix we find.

After we identify the $O(2,2,\IR)$ matrix in question, we are
immediately led to our main purpose, which is to find the
three-parameter deformations of the Sasaki-Einstein manifolds
$T^{1,1}$ and $Y^{p,q}$, whose associated $AdS_5$ geometries are
the near horizon limit of a stack of N D-branes placed at the tip
of their Calabi-Yau cones. The dual field theories were found in
\cite{KW} for $T^{1,1}$ and \cite{MS1,MS2} for $Y^{p,q}$. One
would expect, following \cite{F, beisert, FRT2} that the new
3-parameter deformations we find here should correspond to
non-supersymmetric marginal deformations of the dual field theory.

The plan of the paper is as follows. In the next section, we will
examine the solution generating symmetries used in \cite{LM} and
identify the $O(2,2,\IR)$ matrix, which acts on $E=g+B$. In
section 3, we will reproduce the results of \cite{LM} by the
action of the matrix we find. In section 4, we first discuss the
3-parameter generalization of \cite{F}. Then we present our new
solutions, which are the 3-parameter deformations of the
Sasaki-Einstein manifolds $T^{1,1}$ and $Y^{p,q}$. All our
discussions until section 5 focus on the NS sector of the
solutions, whereas section 5 is devoted to the discussion of the
RR sector. After discussing the regularity of the solutions in
section 6, we conclude with section 7.


\section{A Closer Look at The Solution Generating Symmetries}

Perhaps the best way to study the solution generating
transformations of \cite{LM} is to look at the U-duality group of
the eight dimensional TypeII string theory: $SL(2, \IR) \times
SL(3, \IR).$ Obviously,  we are not being rigorous by calling this
solution generating group the U-duality group, as in fact the
U-duality group is the discrete  $SL(2,Z) \times SL(3,Z)$
\cite{TH}. However, for ease of language, we will refer to the
above group as the U-duality group just as we will, throughout the
text, refer to the solution generating $O(d,d,\IR)$ as the
T-duality group. Let us first examine the subgroup $SO(2,2,\IR)
\simeq SL(2, \IR)_{\tau} \times SL(2,\IR)_{\rho}$, which is the
``trivial" part of the T-duality group $O(2,2,\IR)$ acting within
IIA or IIB\footnote{Obviously, $O(2,2,Z)$ also has matrices with
determinant -1, including the transformation which exchanges the
K\"{a}hler and the complex structure moduli. Sometimes only this
transformation is referred to as being the T-duality
transformation (as in \cite{F}), and sometimes it is referred to
as the non-trivial part of the T-duality group. For an extensive
review of T-duality, see \cite{T-duality1}.}. Here the subindexes
$\tau$ and $\rho$ refers to the fact that the first $SL(2,\IR)$
acts on the complex structure modulus $\tau$ of the internal
2-torus, whereas the second one acts on its K\"{a}hler modulus
$\rho$. $SL(2, \IR)_{\tau}$ is the geometric part of the T-duality
associated with invariance under large diffeomorphisms of the two
torus on which the ten dimensional type IIB string theory has been
compactified. Its action on the complex structure modulus $\tau =
\tau_1 + i \tau_2$ and the background matrix $E = g + B$ is:
\begin{equation}\label{bir}
\tau \longrightarrow \frac{a \tau + b}{c \tau +d}, \ \ \ E
\longrightarrow A E A^t.
\end{equation}
Here $$A = \left(\begin{array}{cc}
  a & b \\
  c & d \\
\end{array}\right)$$ is an $SL(2, \IR)$ matrix  embedded in
$O(2,2, \IR)$ as
\begin{equation}\label{geom}
\left(\begin{array}{cc}
  A & 0_2 \\
  0_2 & (A^t)^{-1} \\
\end{array}\right),
\end{equation}
so that the background matrix $E$ transforms as in (\ref{bir}).

On the other hand, the $SL(2, \IR)_{\rho}$  acts on the K\"{a}hler
modulus $\rho = \rho_1 + i \rho_2 = B_{12} + i \sqrt{g}$ as
\begin{equation}\label{uc}
\rho \longrightarrow \frac{a \rho + b}{c \rho + d}
\end{equation}
with $ad - bc = 1$.  It is important to know how
$SL(2,\IR)_{\rho}$ is embedded into  $O(2,2,\IR)$. One can  see
that the the action of $SL(2,\IR)_{\rho}$ on the background matrix
$E$ is through the following $O(2,2, \IR)$ matrix
\begin{equation}\label{dort}
T = \left(\begin{array}{cc}
  A & B \\
  C & D \\
\end{array}\right) = \left(\begin{array}{cc}
\begin{array}{cc}
     a & \ 0  \\
     0 & \ a  \\
  \end{array} & \begin{array}{cc}
    \ 0 & b \\
    -b \ & 0 \\
  \end{array} \\
  \begin{array}{cc}
   \ 0 & -c \\
   \ c & \ 0 \\
  \end{array} & \begin{array}{cc}
    \ \ d &  \ 0 \\
    \ \ 0 &  \ d  \\
  \end{array} \\
\end{array}\right).
\end{equation}
It is easy to check that as the background matrix transforms as
\cite{T-duality1}
\begin{equation}\label{onemli}
E \longrightarrow  (A E + B)(C E + D)^{-1} \equiv \frac{A E + B}{C
E + D}
\end{equation}
 with $A,B,C,D$ as in (\ref{dort}),  $\rho_1=B_{12}$ and
$\rho_2=\sqrt{g}$ transform as
$$\rho_1 \longrightarrow \frac{ac(\rho_1^2 + \rho_2^2) + (ad+bc)\rho_1
+ bd}{c^2 (\rho_1^2 + \rho_2^2) + 2dc \rho_1 + d^2}$$
$$\rho_2 \longrightarrow \frac{\rho_2}{c^2 (\rho_1^2 + \rho_2^2) + 2dc \rho_1 +
d^2}$$ which is equivalent to (\ref{uc}).

Yet another $SL(2,\IR)$ symmetry of the IIB string theory is the
S-duality, which already exists in ten dimensions and whose $Z_2$
part acts on the string coupling as a strong-weak coupling
duality. This group, which we will denote by $SL(2,\IR)_s$
combines with the non-geometric $SL(2,\IR)_{\rho}$ to form the
$SL(3,\IR)$ part of the U-duality group. From the point of view of
type IIA string theory, this $SL(3,\IR)$ is the geometric part of
the U-duality group, associated with the large diffeomorphisms  of
the three torus on which the eleven dimensional M-theory has been
compactified. On the other hand, the geometric $SL(2,\IR)_{\tau}$
of IIB is mapped under T-duality to a non-geometric $SL(2,\IR)$
which acts on the K\"{a}hler modulus of Type IIA.

The new solutions of \cite{LM} are generated  through the action
of $SL(3,\IR)$. The action of  $SL(2,\IR)_s$ generates new
solutions which correspond to a $\beta$-deformation of the field
theory with a complex parameter $\beta = \sigma$. In this paper,
we will only be interested in real deformations.

The  supergravity solutions of Lunin and Maldacena, which
correspond to  $\beta$-deformations with a real parameter $\beta =
\gamma$ of the dual field theory are obtained by the action of
$SL(2,\IR)_{\rho}$ on $\rho$ as in (\ref{uc}), with the particular
choice $a=d=1, b=0, c=\gamma$ (so that the resulting solutions are
regular). From (\ref{dort}) we see that the relevant T-duality
matrix which acts on the background matrix as in (\ref{onemli}) is
\begin{equation}\label{matrix}
T = \left(\begin{array}{ccc}
  1_2 & & 0_2 \\
 \Gamma & & 1_2 \\
\end{array}\right)
\end{equation} where $1_2$ and $0_2$ are the $2 \times 2$
identity and zero matrices respectively, whereas
\begin{equation}\label{matrixc}
\Gamma = \left(\begin{array}{cc}
  0 & - \gamma \\
  \gamma & \ \ 0 \\
\end{array}\right).
\end{equation}
Perhaps it is useful at this point to recall once again that this
is merely a solution generating action and not a T-duality one. In
fact, as was explained in \cite{LM}, on the field theory side the
corresponding deformation of the field theory superpotential is
$${\rm Tr}(\Phi_1 \Phi_2 \Phi_3 - \Phi_1 \Phi_3 \Phi_2) \rightarrow
{\rm Tr}(e^{i\pi \gamma}\Phi_1 \Phi_2 \Phi_3 - e^{-i\pi
\gamma}\Phi_1 \Phi_3 \Phi_2).$$ When $\gamma$ is  integer,
(\ref{matrix}) becomes an element of the T-duality group
$O(2,2,Z)$ of IIB string theory and the superpotential remains the
same. There is no deformation at all. Of course, this only makes
sense. From the world-sheet point of view, $O(d,d,\IR)$ acts on
the moduli space of 2d CFT's\footnote{When the background is flat,
with the topology of $T^d$, the moduli space is
$O(d,d,\IR)/O(d,\IR) \times O(d,\IR)$. For curved backgrounds with
$d$ commuting isometries, it is more complicated but there is
still a local  $O(d,d,\IR)$ action on the moduli space.} (on the
world-sheet of the string theory whose target space is the
background in question), taking one CFT to another connected to it
by an exactly marginal deformation \cite{T-duality1}. On the field
theory side, this corresponds to starting with the four
dimensional $\CN = 4$ SCFT and deforming it to an $\CN=1$ SCFT
with $\gamma$ parameterizing the deformation (for the particular
example of \cite{LS}). When $\gamma$ is integer, there is no
deformation and the four dimensional conformal field theory
remains the same, just like the discrete $O(d,d,Z)$ maps the
world-sheet CFT to a physically equivalent one.

Let us have a closer look at $T$ in (\ref{matrix}). It can be
factorized in terms of the generators of $O(d,d,\IR)$ in several
different ways. One way is
$$T = g_{D_1} . S \ . g_{D_1}$$ where
$S$ is as in (\ref{geom}) with $$A = \left(\begin{array}{cc}
  1 & \gamma \\
  0 & 1 \\
\end{array}\right)$$ and $g_{Di}$ is the matrix of the $R \rightarrow 1/R$ transformation, called
the ``factorized duality" in \cite{T-duality1}, along the $i$th
direction:
$$\left(\begin{array}{cc}
  1_2 - e_i & e_i \\
  e_i & 1_2-e_i \\
\end{array}\right),$$ $e_i$ being the $2\times2$ matrix with all
entries 0 except $ii$th entry, which is 1. So, $T$ corresponds to
a factorized  duality along one leg of the 2-torus, an $SL(2,\IR)$
transformation and
 another factorized duality along the same leg. This is the
interpretation that was used in \cite{F}. On the other hand, $T$
can also be factorized as
\begin{equation}\label{factor}
T = \left(\begin{array}{cc}
  0 & 1_2 \\
  1_2 & 0 \\
\end{array}\right) . \left(\begin{array}{cc}
  1_2 & \Gamma \\
  0 & 1_2 \\
\end{array}\right) . \left(\begin{array}{cc}
  0 & 1_2 \\
  1_2 & 0 \\
\end{array}\right).
\end{equation}
This is a volume inversing factorized duality along both legs of
the torus, a geometric shift of the B-field, and then  dualizing
back along the two legs. This factorization  will generalize to
the 3-parameter deformations, as we will see in section 4.

For future use, let us note how the momenta $p^i$ and the winding
charges $w^i$, $i=1,\cdots,d, \ d={\rm dim} T^d$ associated with
the world-sheet currents $\partial_t x^i$ and $\partial_s x^i$
transform under the action of a T-duality matrix of the type
(\ref{matrix}). Let $Z$ be the $2d$-dimensional vector
\begin{equation}
Z = \left(
\begin{array}{c}
  w^i \\
  p^i \\
\end{array}\right).
\end{equation}
Under the action of a generic  $T \in O(d,d,\IR)$, it transforms
as \cite{T-duality1}
\begin{equation}\label{winding}
Z \longrightarrow (T^t)^{-1} Z.
\end{equation}
In our case, in which $d=2$ and $T$ is of the form (\ref{matrix}),
we have
\begin{equation}\label{winding}
\left(\begin{array}{c}
  w^1 \\
  w^2 \\
  p^1 \\
  p^2 \\
\end{array}\right) \longrightarrow \left(\begin{array}{c}
  w^1 - \gamma p^2 \\
  w^2 + \gamma p^1 \\
  p^1 \\
  p^2 \\
\end{array}\right).
\end{equation}


\section{A Shortcut to the Lunin-Maldacena Backgrounds}
Now that we have the T-duality matrix whose action generates the
new solutions we can  reproduce the Lunin-Maldacena backgrounds in
an easier way. Below we will demonstrate this briefly so that our
discussions for the generalized case will be more transparent. In
this section, we will only be concerned with the NS sector,
leaving the transformation of the RR fields to section 5. We are
leaving the discussion of $AdS_5 \times S^5$ to the end, as it
involves an extra feature, which will lead us to our new
3-parameter deformations.

\subsection{$ AdS_5 \times T^{1,1}$}

Our first example is $AdS_5 \times T^{1,1}$, where $T^{1,1}$ is
the coset space $(SU(2) \times SU(2))/U(1)$. This is the dual of
${\cal{N}} = 1$ supersymmetric Yang-Mills theory, which arises
from a stack of N D-branes at the tip of the conifold, the conic
Calabi-Yau 3-fold whose base space is $T^{1,1}$ \cite{KW}. The
metric of $AdS_5 \times T^{1,1}$ is
\begin{equation}\label{t11}
\frac{ds^2}{R^2} = ds_{AdS}^2 + \frac{1}{6}
\sum_{i=1}^{2}(d\theta_i^2 + \sin^2\theta_i d\phi_i^2) +
\frac{1}{9}(d\psi + \cos\theta_1 d\phi_1 + \cos\theta_2
d\phi_2)^2.
\end{equation}
This is clearly an $S^1$ bundle over $S^2 \times S^2$ with $\psi$
parameterizing the fiber circle which winds over the two base
spheres once\footnote{When the Hopf numbers of the $S^1$ bundle
over the base 2-spheres are $p$ and $q$ the above generalizes to
the manifolds $T^{p,q}$.}. The isometry group of $T^{1,1}$ is
$SU(2) \times SU(2) \times U(1)$ and in particular, it has three
commuting Killing vectors: $\partial/\partial \phi_1,
\partial/\partial \phi_2, \partial /
\partial \psi$. \cite{LM} used the first two of them in order to
generate new solutions corresponding to exactly marginal real
$\gamma$-deformations of the field theory. Now we will see how the
action of (\ref{matrix}) does indeed reproduce their result.

First note that as $S^1$ is nontrivially fibered over the two
isometry directions $\phi_1 $ and $ \phi_2 $, we cannot simply use
the matrix in (\ref{matrix}) but rather should embed it in
$O(3,3,\IR)$. The rules of this were given in \cite{T-duality2}
and the result is
\begin{equation}\label{matrix2}
T = \left(\begin{array}{ccc}
 1_3 & & 0_3 \\
  \Gamma & & 1_3 \\
\end{array}\right)
\end{equation}
 with $1_3$
and $0_3$  being $3 \times 3$ identity and zero matrices and
$\Gamma$ becomes:
\begin{equation}\label{matrixyeni}
\Gamma = \left(\begin{array}{ccc}
  0 & -\gamma & 0 \\
  \gamma & 0 & 0 \\
  0 & 0 & 0 \\
\end{array}\right).
\end{equation}
Applying this to the background matrix\footnote{Here we use the
notation of \cite{LM}: $c_i = \cos \theta_i$ and $s_i = \sin
\theta_i$.}
\begin{equation}\label{t11e}
E = R^2\left(\begin{array}{ccc}
  \frac{c_1^2}{9}+\frac{s_1^2}{6} & \frac{c_1 c_2}{9} & \frac{c_1}{9} \\
 \frac{c_1 c_2}{9}  & \frac{c_2^2}{9}+\frac{s_2^2}{6} & \frac{c_2}{9} \\
  \frac{c_1}{9} & \frac{c_2}{9} & \frac{1}{9} \\
\end{array}\right)
\end{equation}
we obtain the new background matrix $E' = g' + B' = E .(\Gamma E +
1_3)^{-1}$ from which we read
\begin{equation}\label{t11b}
\frac{ds^2}{R^2} = ds_{AdS}^2 + G[\frac{1}{6}
\sum_{i=1}^{2}(G^{-1}d\theta_i^2 + \sin^2\theta_i d\phi_i^2) +
\frac{1}{9}(d\psi + \cos\theta_1 d\phi_1 + \cos\theta_2 d\phi_2)^2
+ \hat{\gamma}^2 \frac{s_1^2 s_2^2}{324} d\psi^2].
\end{equation}
\begin{equation}\label{t11b}
\frac{B}{R^2} = \hat{\gamma} G [(\frac{s_1^2 s_2^2}{36} +
\frac{c_1^2 s_2^2 + c_2^2 s_1^2}{54})d\phi_1 \wedge d\phi_2 +
\frac{s_1^2 c_2}{54} d\phi_1 \wedge d\psi - \frac{c_1 s_2^2}{54}
d\phi_2 \wedge d\psi]
\end{equation}
Here $$G = {\rm det}(\Gamma E + 1_3)^{-1} = (1+
\hat{\gamma}^2(\frac{c_1^2 s_2^2+c_2^2 s_1^2}{54} + \frac{s_1^2
s_2^2}{36}))^{-1}, \ \ \ \ \hat{\gamma} = R^2 \gamma .$$ The
dilaton always transforms as\footnote{We want the asymptotic value
$\Phi_0$ of the dilaton  to have the same value as in the $AdS_5
\times X_5$ case, as the exactly marginal operator by which the
field theory is deformed does not couple to the dilaton. As a
result, $R = (4 \pi e^{\Phi_0} N)^{1/4}$ of the deformed
background is the same as that of the original background.}
\cite{T-duality1}
\begin{equation}\label{dilaton} e^{2\Phi} \longrightarrow
\frac{e^{2\Phi}}{{\rm det}(\Gamma E + 1_3)}
\end{equation}
so we have $e^{2 \Phi '} = G e^{2 \Phi}.$ This is the exact same
background as in \cite{LM}, although presented in a different
way\footnote{Except for an overall $G^{-1/4} = e^{-\Phi
'/2}/e^{-\Phi /2}$ factor, as we are presenting our results in the
string frame unlike in \cite{LM}, where it is presented in the
Einstein frame.}.


\subsection{$ AdS_5 \times Y^{p,q}$}

The next backgrounds we will consider are the recently discovered
Sasaki-Einstein manifolds $Y^{p,q}$ \cite{MSJ,MS1}. The dual field
theory arises from a stack of N  D-branes placed at the tip of the
toric Calabi-Yau cone over $Y^{p,q}$ \cite{MS1,MS2}. The metric is
\begin{eqnarray}\label{SE}
\frac{ds^2}{R^2} &=& ds^2_{AdS_5} +
\frac{1-y}{6}(d\theta^2+s_{\theta}^2d\phi^2) +
\frac{1}{w(y)q(y)}dy^2 + \frac{q(y)}{9}(d\psi-c_{\theta}d\phi)^2
\\
& & + w(y)[l d\alpha + f(y)(d\psi - c_{\theta}d\phi)]^2
\end{eqnarray}Here $$w(y) = \frac{2(a-y^2)}{1-y}, \ \ \ q(y) =
\frac{a-3y^2+2y^3}{a-y^2}, \ \ \ f(y) =
\frac{a-2y+y^2}{6(a-y^2)}$$ $$l = \frac{q}{3q^2 -
2p^2+p(4p^2-3q^2)^{1/2}}.$$ So the background matrix $E = g+b = g$
is
\begin{equation}\label{bgSE}
E = R^2\left(\begin{array}{ccc}
  g_{11} & g_{12} & g_{13} \\
  g_{12} & g_{22} & g_{23} \\
  g_{13} & g_{23} & g_{33} \\
\end{array}\right),
\end{equation}
where
\begin{eqnarray}\label{contents}
g_{11} &=& l^2 w(y), \ \  g_{22} = \frac{1-y}{6} s_{\theta}^2 +
\frac{q(y)}{9} c_{\theta}^2 + w(y)f(y)^2c_{\theta}^2, \ \  g_{12}
= -l w(y) f(y) c_{\theta} \nonumber \\
g_{33} &=& \frac{q(y)}{9} + w(y)f(y)^2, \ \   g_{13} = l w(y)
f(y), \ \ g_{23} = -\frac{q(y)}{9}c_{\theta} - w(y)f(y)^2
c_{\theta}.
\end{eqnarray}
We find the new background metric and the B-field from the
symmetric and the antisymmetric parts of $E' = E (\Gamma E +
1_3)^{-1}$, where $\Gamma$ is as in (\ref{matrixyeni}). First let
us write the determinant of $G = {\rm det}(\Gamma E + 1_3)^{-1}$.
$$G = \frac{1}{1 + \hat{\gamma}^2 \Delta},$$
where
\begin{equation}\label{delta3}
\Delta=\frac{2q(y)c_{\theta}^2+3(1-y)s_{\theta}^2}{9(1-y)}(a-y^2)l^2,
\ \ \ \ \ \ \hat{\gamma} = R^2 \gamma .
\end{equation}
As mentioned in the previous subsection the dilaton will simply
transform as $e^{2 \Phi} \rightarrow G e^{2 \Phi}$. Doing the
matrix multiplication we also find
\begin{eqnarray}\label{genel}
g_{ij} & \longrightarrow & G g_{ij}, \ \ \ \ \ {\rm except} \ g_{33} \\
g_{33} & \longrightarrow & G (g_{33} + \hat{\gamma}^2 e)
\end{eqnarray}
where $e$ is the determinant of $E$ in (\ref{bgSE})
\begin{eqnarray}\label{det}
e = \frac{{\rm det}E}{R^6} &=& g_{33}(g_{11}g_{22} - g_{12}^2) +
2g_{12}g_{13}g_{23} - g_{13}^2g_{22} - g_{23}^2 g_{11} \nonumber \\
&=& \frac{1}{27} l^2 (2 y^3 - 3 y^2 + a) s_{\theta}^2.
\end{eqnarray}
On the other hand we find $$\frac{B}{R^2} = B_{12} d\alpha \wedge
d\phi + B_{13} d\alpha \wedge d\psi + B_{23} d\phi \wedge d\psi$$
where
\begin{eqnarray}\label{genelb}
B_{12} &=& \hat{\gamma}(g_{11}g_{22} - (g_{12})^2)= \hat{\gamma}
\Delta, \nonumber \\
B_{13} &=& \hat{\gamma}(g_{11}g_{23} - g_{12}g_{13})= \hat{\gamma}
\frac{2l^2(a+2y^3-3y^2)c_{\theta}}{9(y-1)} \nonumber \\
B_{23} &=& \hat{\gamma}(g_{12}g_{23} - g_{22}g_{13}) = -
\hat{\gamma} \frac{l(a-2y+y^2)s_{\theta}^2}{18}.
\end{eqnarray} One can check that the new background is exactly
the same as that presented in \cite{LM}.


\subsection{$ AdS_5 \times S^5 $}

Finally we consider the background $AdS_5 \times S^5$ whose metric
is
\begin{equation}\label{s5}
\frac{ds^2}{R^2} = ds_{AdS_5}^2 + \sum_{i=1}^{3} d\mu_i^2 +
\mu_i^2 d\phi_i^2, \ \ \ \ \ {\rm with} \ \ \ \sum_{i} \mu_i^2 =
1.
\end{equation}
When we perform the T-duality transformation with the matrix
(\ref{matrix2}) we see that the resulting background is not the
same as in \cite{LM}. After a moment's thought, one realizes that
the T-duality matrix, which generates the deformed solution in
\cite{LM} is in fact
\begin{equation}\label{matrix3}
T = \left(\begin{array}{ccc}
  1_3 &  & 0_3 \\
  \Gamma &  & 1_3 \\
\end{array}\right)
\end{equation}
where
\begin{equation}\label{gamma3}
\Gamma = \left(\begin{array}{ccc}
  0 & -\gamma & \gamma \\
  \gamma & 0 & -\gamma \\
  -\gamma & \gamma & 0 \\
\end{array}\right)
\end{equation}
Of course, first doing a coordinate transformation, then a
T-duality of the form (\ref{matrix2}) and an inverse coordinate
transformation to the original coordinates will give a T-duality
transformation whose matrix is
\begin{equation}\label{trmatrix}
\left(\begin{array}{ccc}
  1_3 &  & 0_3 \\
  A^t \Gamma A &  & 1_3 \\
\end{array}\right).
\end{equation}
So, with an appropriate choice of coordinate transformations, one
can obtain a T-duality matrix of the form (\ref{matrix3}),
starting from (\ref{matrix2}). This is what was done in \cite{LM},
see for example \cite{F}.

\noindent Applying (\ref{matrix3}) one obtains the background in
\cite{LM}:
\begin{eqnarray}\label{news5}
ds^2 & = & R^2[ds^2_{AdS_5} + \sum_i(d\mu_i^2 + G \mu_i^2
d\phi_i^2) + \hat{\gamma}^2 G \mu_1^2 \mu_2^2 \mu_3^2(\sum_id\phi_i)^2] \\
B \ & = & \hat{\gamma} R^2 G(\mu_1^2 \mu_2^2 d\phi_1 \wedge
d\phi_2 + \mu_2^2 \mu_3^2 d\phi_2 \wedge d\phi_3 + \mu_3^2 \mu_1^2
d\phi_3 \wedge d\phi_1) \\
e^{2 \Phi '}& = &G e^{2 \Phi} \\
G \ & = & (1 + \hat{\gamma}^2(\mu_1^2 \mu_2^2 + \mu_2^2 \mu_3^3 +
\mu_1^2 \mu_3^2))^{-1}, \ \ \ \ \ \hat{\gamma} = R^2 \gamma
\end{eqnarray}
At this point, one can readily guess how the generalizations with
three parameters will arise.


\section{Three Parameter Generalizations}
\subsection{Three-parameter deformation of $AdS_5 \times S^5$: Frolov's solution}
As we discussed above, deformations of the $AdS_5 \times S^5$
background can be obtained by transforming with the T-duality
matrix (\ref{matrix3}), which is also equivalent to performing a
coordinate transformation, a  factorized duality along one
isometry direction, a geometric shift, then dualizing back along
the same isometry direction and finally the inverse coordinate
transformation. Frolov applied the latter on all the 2-tori
embedded in the 3-torus and found a 3-parameter family of
deformations of the $AdS_5 \times S^5$ background \cite{F}. As one
might guess from our previous discussions, it is equivalent to
generating a new solution by the action of the T-duality matrix
\begin{equation}\label{frolovmatrix}
\left(\begin{array}{ccc}
  1_3 &  & 0_3 \\
  \Gamma &  & 1_3 \\
\end{array}\right),
\end{equation}
where $\Gamma $ is now\footnote{Very recently, $d(d-1)/2$
parameter deformations of a background with $d$ commuting $U(1)$
isometries was studied in \cite{frolovnew}. We would expect that
such deformed solutions should be generated by the action of an
$O(d,d,\IR)$ matrix of the same form, where $\Gamma$ is now a
$d$-dimensional antisymmetric matrix with $d(d-1)/2$ independent
parameters.}
\begin{equation}\label{frolovgamma}
\left(\begin{array}{ccc}
  0 & -\gamma_3 & \gamma_2 \\
  \gamma_3 & 0 & -\gamma_1 \\
  -\gamma_2 & \gamma_1 & 0 \\
\end{array}\right).
\end{equation}

Applying the T-duality matrix (\ref{frolovmatrix}) to the
background matrix of (\ref{s5}) we obtain exactly the solution of
Frolov, whose NS sector is:
\begin{eqnarray}\label{3s5}
ds^2 & = & R^2[ds^2_{AdS_5} + \sum_i(d\mu_i^2 + G \mu_i^2
d\phi_i^2) +  G \mu_1^2 \mu_2^2 \mu_3^2(\sum_i \hat{\gamma}_i d\phi_i)^2] \nonumber \\
B \ & = & R^2 G(\hat{\gamma}_3 \mu_1^2 \mu_2^2 d\phi_1 \wedge
d\phi_2 +\hat{\gamma}_1  \mu_2^2 \mu_3^2 d\phi_2 \wedge d\phi_3 +
\hat{\gamma}_2 \mu_3^2 \mu_1^2
d\phi_3 \wedge d\phi_1) \nonumber \\
e^{2 \Phi '}& = &G e^{2 \Phi} \nonumber \\
G \ & = & (1 + (\hat{\gamma}_3^2 \mu_1^2 \mu_2^2 + \hat{\gamma}_1
\mu_2^2 \mu_3^3 + \hat{\gamma}_2 \mu_1^2 \mu_3^2))^{-1}, \ \ \ \ \
\hat{\gamma}_i = R^2 \gamma_i.
\end{eqnarray}

\noindent Before we move on to obtain the 3-parameter deformations
of the toric backgrounds with the same method, let us pause for a
moment and look at the solution generating matrix
(\ref{frolovmatrix}) more carefully. First of all, as the
$\gamma_i$ are now independent, the only possible factorization of
(\ref{frolovmatrix}) in terms of the generators of $O(d,d,\IR)$ is
\begin{equation}\label{factor2}
\left(\begin{array}{ccc}
  1_3 &  & 0_3 \\
  \Gamma &  & 1_3 \\
\end{array}\right) = \left(\begin{array}{cc}
  0 & 1_3 \\
  1_3 & 0 \\
\end{array}\right) . \left(\begin{array}{ccc}
  1_3 &  & \Gamma \\
  0 &  & 1_3 \\
\end{array}\right) . \left(\begin{array}{cc}
  0 & 1_3 \\
  1_3 & 0 \\
\end{array}\right).
\end{equation}
So, the action of (\ref{frolovmatrix}) is equivalent to a
factorized duality along all the isometry directions, then doing a
$\Theta$ shift\footnote{It is standard to call the generators of
$O(d,d,\IR)$ of this type $\Theta$ shifts \cite{T-duality1}.}
 and finally an inverse factorized duality along all the
isometry directions. So, perhaps it is convenient, in the spirit
of \cite{F},  to call this a T$\Theta$T transformation.

We also would  like to discuss the transformation of the vector of
momentum and winding charges, as this played a crucial role in
\cite{F}. Using (\ref{winding}) we see that the momenta are
conserved and if $w_i$ are the winding numbers of the undeformed
background then the deformed background has
\begin{eqnarray}\label{twisted}
\phi_1(2\pi) - \phi_1(0) &=& 2\pi(w_1 + \gamma_2 J_3 - \gamma_3 J_2) \\
\phi_2(2\pi) - \phi_2(0) &=& 2\pi(w_2 + \gamma_3 J_1 - \gamma_1 J_3) \\
\phi_3(2\pi) - \phi_3(0) &=& 2\pi(w_3 + \gamma_1 J_2 -\gamma_2
J_1).
\end{eqnarray}
Here $(J_1, J_2, J_3)$ are the conserved angular
momenta\footnote{Here we use the letter $J$, not $p$, following
\cite{F}, where $p$ is used to name the world-sheet current.}.
This is the generalization of what was used in \cite{F} in order
to write down the Lax pair of the (1-parameter) deformation of the
$AdS_5 \times S^5$ background. Another crucial information that
\cite{F} needed to write the Lax pair was the transformation of
coordinates.  Under a general T-duality transformation
(\ref{onemli}), the metric always transforms as
\begin{equation}\label{onemli2}
g \longrightarrow K^t g K
\end{equation}
where $$K = (C E + D)^{-1}.$$ Instead of this ``active
transformation" one can also consider the ``passive
transformation" of the coordinates
\begin{equation}\label{onemli3}
d\phi_i \longrightarrow d\tilde{\phi}_i = K \ d\phi_i
\end{equation}
so that the line element $d\phi^t  g  d\phi$  still transforms in
the same way. (\ref{onemli3}) gives the following on-shell
relations between the two coordinate systems
\begin{eqnarray}\label{onshell}
d \tilde{\phi}_1 & = & G[(1+\hat{\gamma}_1^2 \mu_2^2
\mu_3^2)d\phi_1 + (\hat{\gamma}_3 \mu_2^2 + \hat{\gamma}_1
\hat{\gamma}_2 \mu_2^2 \mu_3^2)d\phi_2 - (\hat{\gamma}_2
\mu_3^2 - \hat{\gamma}_1 \hat{\gamma}_3 \mu_2^2 \mu_3^2)d\phi_3] \nonumber \\
d \tilde{\phi}_2 & = & G[-(\hat{\gamma}_3 \mu_1^2 - \hat{\gamma}_1
\hat{\gamma}_2 \mu_1^2 \mu_3^2)d\phi_1 + (1+\hat{\gamma}_2^2
\mu_1^2 \mu_3^2)d\phi_2 \nonumber
+ (\hat{\gamma}_1 \mu_3^2 + \hat{\gamma}_2 \hat{\gamma}_3 \mu_1^2 \mu_3^2)d\phi_3] \\
d \tilde{\phi}_3 & = & G[(\hat{\gamma}_2 \mu_1^2 + \hat{\gamma}_1
\hat{\gamma}_3 \mu_1^2 \mu_2^2)d\phi_1 - (\hat{\gamma}_1 \mu_2^2 -
\hat{\gamma}_2 \hat{\gamma}_3 \mu_1^2 \mu_2^2)d\phi_2
+(1+\hat{\gamma}_3^2 \mu_1^2 \mu_2^2)d\phi_3]
\end{eqnarray}
where $\hat{\gamma}_i$ and $G$ are as in (\ref{3s5}). This played
a crucial role in \cite{F} in writing the Lax pair for the strings
in the deformed background.


\subsection{Three-parameter deformation of $T^{1,1}$ and $Y^{p,q}$: New solutions}

We are finally in a position to write our  new three-parameter
deformations of the Sasaki-Einstein manifolds $T^{1,1}$ and
$Y^{p,q}$. All we have to do is to generate new solutions by the
action of the matrix (\ref{frolovmatrix}). Following \cite{F}, one
would expect our new solutions to correspond to non-supersymmetric
marginal deformations of the dual field theory\footnote{In
\cite{russo}, it was shown that the spectrum of string theory in
the deformed (with three parameters) flat space contains tachyons.
It was argued in \cite{frolovnew} that this does not necessarily
imply the unstability of the string theory on the deformed $AdS_5
\times S^5$.}. For the 3-parameter deformation of $AdS_5 \times
S^5$, the corresponding integrable spin chain   and the Bethe
ansatz were constructed in \cite{beisert}. Further checks of the
AdS/CFT correspondence  were performed in \cite{FRT2}. This
motivates us to study the three-parameter deformations of the
toric backgrounds, even though the resulting theory will be
non-supersymmetric. Note that, as the form of the solution
generating matrix is the same, (\ref{twisted}) and (\ref{onemli3})
will also remain the same. Obviously, the matrix $K$ depends on
the background matrix $E$, therefore it will be more complicated
in our case. Before we move on to the next point, it would be
relevant to make a remark about the supersymmetry, or the lack
thereof. Let's have a look back at Frolov's solution. We do not
expect this solution to be supersymmetric because we have
T-dualized along a $U(1)$ direction with respect to which the
Killing spinors have non-vanishing Lie derivative. However, a
curious thing happens when all $\gamma_i$ are integers. As was
discussed before, in this case there is no deformation to the
field theory, so the dual field theory is supersymmetric. Perhaps,
this is not so curious, when one remembers the existence of
supersymmetric string vacua for which the corresponding
supergravity solution does not have any Killing spinors. This is
the phenomena of ``supersymmetry without supersymmetry", which was
discussed in \cite{duff,hullholonomy}. After this remark, we
simply present the new backgrounds below.

~

\underline{\textit{$AdS_5 \times T^{1,1}$}}:
\begin{eqnarray}\label{newt11}
\frac{ds^2}{R^2} &=& ds_{AdS_5}^2 + G[\frac{1}{6}
\sum_{i=1}^{2}(G^{-1}d\theta_i^2 + s_i^2 d\phi_i^2) +
\frac{1}{9}(d\psi
+ c_1 d\phi_1 + c_2 d\phi_2)^2 +  \\
& & + \frac{s_1^2 s_2^2}{324} (\hat{\gamma}_3 d\psi +
\hat{\gamma}_1 d\phi_1 + \hat{\gamma}_2 d\phi_2)^2] \nonumber \\
\frac{B}{R^2} &=& G[(\hat{\gamma}_3(\frac{c_1^2 s_2^2 + c_2^2
s_1^2}{54} + \frac{s_1^2 s_2^2}{36})-\hat{\gamma}_2
\frac{c_2s_1^2}{54} - \hat{\gamma}_1 \frac{c_1s_2^2}{54}) d\phi_1
\wedge
d\phi_2 \nonumber \\
& & + \frac{\hat{\gamma}_3 s_1^2 c_2 - \hat{\gamma}_2 s_1^2}{54}
d\phi_1 \wedge d\psi - \frac{\hat{\gamma}_3 c_1 s_2^2 -
\hat{\gamma}_1 s_2^2}{54}
d\phi_2 \wedge d\psi \\
\ e^{2 \Phi '} & = & G e^{2 \Phi}, \ \ \ \ \ \ \ \ \ \ \  \ \ \ \ \ \   \ \ \hat{\gamma}_i = R^2 \gamma_i \nonumber \\
G &=& (1+(\hat{\gamma}_3^2(\frac{c_1^2 s_2^2 + c_2^2 s_1^2}{54} +
\frac{s_1^2 s_2^2}{36}) + \hat{\gamma}_2^2 \frac{s_1^2}{54} +
\hat{\gamma}_1^2 \frac{s_2^2}{54} - \hat{\gamma}_3 \hat{\gamma}_2
\frac{s_1^2 c_2}{27} - \hat{\gamma}_1 \hat{\gamma}_3 \frac{c_1
s_2^2}{27}))^{-1} \nonumber
\end{eqnarray}

~

\noindent This solution reduces to that of \cite{LM} when
$\gamma_3 = \gamma$ and $\gamma_1 = \gamma_2 = 0$.

~

\underline{\textit{$AdS_5 \times Y^{p,q}$}}:

~

\noindent The background obtained by applying the solution
generating matrix (\ref{frolovmatrix}) has the metric components
\begin{equation}\label{100}
g_{ij} \longrightarrow G (g_{ij} + \hat{\gamma}_i \hat{\gamma}_j
e),
\end{equation}
where $e$ is as in (\ref{det}) and
\begin{equation}\label{G}
G^{-1} = 1 + \sum_i \hat{\gamma}_i^2 \Delta_i + 2 \hat{\gamma}_1
\hat{\gamma}_2 K_3 + 2\hat{\gamma}_1 \hat{\gamma}_3 K_2 + 2
\hat{\gamma}_2 \hat{\gamma}_3 K_1
\end{equation}
with
\begin{equation}
K_1  = g_{12}g_{13} -g_{11}g_{23}, \ \ \ \ \ K_2  =  g_{12}g_{23}
- g_{22}g_{13}, \ \ \ \  K_3  = g_{13}g_{23}-g_{12}g_{33},
\end{equation}
\begin{equation}
\Delta_1  = g_{22}g_{33} - (g_{23})^2, \ \ \ \ \ \Delta_2  =
g_{11}g_{33} - (g_{13})^2, \ \ \ \ \Delta_3  =  g_{11}g_{22} -
(g_{12})^2.
\end{equation}
$\Delta_3 = \Delta$ in (\ref{delta3}) and $K_1, K_2$ were given in
(\ref{genelb}). On the other hand, $K_3 = 0$ and
\begin{eqnarray}
\Delta_1 & = & \frac{(1-y)s_{\theta}^2}{6}\left(\frac{q(y)}{9}+w(y)f(y)^2\right) = \frac{(2+a-6y+3y^2)s_{\theta}^2}{108}, \nonumber \\
\Delta_2 & = & \frac{l^2 w(y) q(y)}{9} =
\frac{2l^2(a-3y^2+2y^3)}{9(1-y)}.
\end{eqnarray}

\noindent The metric, B-field and the dilaton of the new solution
are
\begin{eqnarray}\label{SEyeni}
\frac{ds^2}{R^2} &=& G[ds^2_{AdS_5} +
\frac{1-y}{6}(d\theta^2+s_{\theta}^2d\phi^2)+
\frac{1}{w(y)q(y)}dy^2 + \frac{q(y)}{9}(d\psi-c_{\theta}d\phi)^2 \\
& & + w(y)[l d\alpha + f(y)(d\psi - c_{\theta}d\phi)]^2 \nonumber
+ \frac{l^2 (2 y^3 - 3 y^2 + a) s_{\theta}^2}{27} (\hat{\gamma}_1
d\alpha + \hat{\gamma}_2 d\phi + \hat{\gamma}_3 d\psi)^2], \nonumber \\
\frac{B}{R^2} & = &  G (B_{12} d\alpha \wedge d\phi +
B_{13} d\alpha \wedge d\psi + B_{23} d\phi \wedge d\psi), \\
\ e^{2 \Phi '} &=& G e^{2 \Phi}, \ \ \ \ \ \ \ \  \ \hat{\gamma}_i
= R^2 \gamma_i,
\end{eqnarray}
where
\begin{eqnarray}\label{BB}
B_{12} & = & \hat{\gamma}_3 \Delta_3 + \hat{\gamma}_1 K_2 +
\hat{\gamma}_2 K_1,
\nonumber \\
 B_{31} & = & \hat{\gamma}_2 \Delta_2 + \hat{\gamma}_1 K_3 +
\hat{\gamma}_3 K_1, \nonumber \\
B_{23} & = & \hat{\gamma}_1 \Delta_1 + \hat{\gamma}_3 K_2 +
\hat{\gamma}_2 K_3.
\end{eqnarray}
As before, this solution reduces to that of \cite{LM} when
$\gamma_3 = \gamma$ and $\gamma_1 = \gamma_2 = 0$.


\section{Ramond-Ramond Fields}

So far, we have only considered the NS sector of the new
solutions. In section 3, we showed that the action of the solution
generating matrix (\ref{matrix2}) and (\ref{matrix3}) generates
the NS sector of the Lunin-Maldacena backgrounds, which
immediately ensures that the RR sectors also agree. Then in
section 4, we obtained new 3-parameter deformations of the toric
backgrounds. The aim of the present section is to study the RR
sector of these new solutions.

Given a T-duality matrix, which acts on the NS sector as in
(\ref{onemli}) one can also work out the transformation of the RR
fields under the action of this matrix. This was given in
\cite{hullb,cvetic,hassan,oota}. Here we will follow \cite{oota},
which focuses on the T-duality transformations acting within Type
IIA or Type IIB, namely those in $SO(d,d)$ with determinant 1.
Surely, our matrix of interest, (\ref{frolovmatrix}) is of this
type.

As was already realized in \cite{TH,witten} RR fields combine with
the NS 2-form field in an appropriate way to transform under the
chiral spinor representation of $O(d,d)$. The details were worked
out in \cite{oota}. Here we give a brief review of their results,
without presenting the details.

The first step is to combine the RR potentials $C_p$ of the Type
II string theory (even forms in IIB and odd forms in IIA) with the
NS 2-form field $B_2$ in the following way:
\begin{eqnarray}\label{RR1}
D_0& \equiv & C_0, \ \ \ \ \ \ \ \ \ \ \ \ \ \ \ \ \ \ \ \ \ \ D_1 \equiv C_1, \\
D_2& \equiv & C_2 + B_2 \wedge D_0, \ \ \ \ \ \ \ D_3 \equiv C_3 +
B_2
\wedge C_1, \nonumber \\
D_4& \equiv & C_4 + \frac{1}{2} B_2 \wedge C_2 + \frac{1}{2} B_2
\wedge B_2 \wedge C_0. \nonumber
\end{eqnarray}
Now introduce
\begin{equation}\label{RR2}
D \equiv \sum_{p=0}^8 D_p, \ \ \ \ \ F \equiv e^{-B_2}
\sum_{p=0}^8 dD_p = \sum_{p=0}^8 F_{p+1}.
\end{equation}
The indices run from 0 to 8, as we have also included the
electromagnetic  duals of the gauge potentials $D_p$\footnote{The
electromagnetic duals $ D_{8-p}$  of $D_p$ are the potential
fields obtained by solving the field equations for the latter.
This ensures that $F$ defined as above satisfies $F_{10-p} =
(-1)^{[\frac{p-1}{2}]} * F_p$ where $[\frac{p-1}{2}]$ is the first
integer greater than or equal to $\frac{p-1}{2}$ \cite{ddII}.}.

Next we introduce $2d$ fermionic operators $\psi_i$ and $\psi^{i
\dagger}$ satisfying
\begin{equation}\label{RR3}
\{\psi_i, \psi^{j \dagger}\} = \delta_i^{\ j}{\bf 1}, \ \ \ \
\{\psi_i, \psi_j\} = \{\psi^{i \dagger}, \psi^{j \dagger}\} = 0, \
\ \ \ (i,j = 1, \cdots, d)
\end{equation}
with $$(\psi_i)^{\dagger} = \psi^{i \dagger}.$$ We construct a
$2^d$ dimensional Fock space spanned by
\begin{equation}\label{RR4}
\mid \alpha > = \psi^{i_1 \dagger} \cdots \psi^{i_n \dagger} \mid
0 > \ \ \ \ \ \ (n = 0,\cdots,d)
\end{equation}
where we have introduced the vacuum $\mid 0 >$ such that $\psi_i
\mid 0>  = 0$ and $<0 \mid 0> = 1$. $\alpha$ in (\ref{RR4}) is a
multi-index $\alpha = (i_1, \cdots, i_n) \ {\rm with} \ (i_1 <
\cdots < i_n)$. Now we would like to construct states in this Fock
space corresponding to the form fields $D$ and $F$ in (\ref{RR2}).
To this end, \cite{oota} utilized the following one-to-one
correspondence between the set of differential forms and the space
of creation operators $\psi^{i \dagger}$ under which a
differential form $\Omega$
$$\Omega = \sum_n \frac{1}{n!} \Omega_{i_1 \cdots i_n} dy^{i_1}
\wedge \cdots \wedge dy^{i_n} = \sum_q \sum_n \frac{1}{n!}
\Omega^{(q)}_{i_1 \cdots i_n} dy^{i_1} \wedge \cdots \wedge
dy^{i_n}$$ is mapped to the following operator
$${\bf \Omega} \equiv \sum_n \frac{1}{n!} \Omega_{i_1 \cdots i_n}
\psi^{i_1 \dagger} \cdots \psi^{i_n \dagger} = \sum_q \sum_n
\frac{1}{n!} \Omega^{(q)}_{i_1 \cdots i_n} \psi^{i_1 \dagger}
\cdots \psi^{i_n \dagger}.$$ On the right hand side, the index
$(q)$ indicates that the degree of $\Omega$ is $q$ as a
differential form on the non-compact space, whereas $i_1, \cdots,
i_n$ are compact indices. This actually gives an isomorphism as an
algebra. Now, to each differential form $\Omega$ one can construct
the following state
\begin{equation}\label{RR5}
\mid \Omega > \equiv {\bf \Omega} \mid 0>.
\end{equation}
The main result of \cite{oota} is that the states corresponding to
$D$ and $F$ in (\ref{RR2})  transform under $SO(d,d)$ as
\begin{equation}\label{RRasil}
\mid F > \longrightarrow {\bf \Lambda} \mid F >, \ \ \ \ \ \ \
\mid D > \longrightarrow {\bf \Lambda} \mid D >
\end{equation}
where $${\bf \Lambda} \mid \beta > = \sum_{\alpha} \mid \alpha >
S_{\alpha \beta}(\Lambda)$$ and $S(\Lambda) = (S_{\alpha
\beta}(\Lambda))$ is the spinor representation.  Here $\Lambda$ is
a given $SO(d,d)$ matrix. \cite{oota} gave the explicit
construction of the operators ${\bf \Lambda}$ corresponding to the
generators of $SO(d,d)$.

Now, let us turn to our case of interest where $d=3$ and the
$SO(3,3)$ matrix of concern is either of the form (\ref{matrix2}),
(\ref{matrix3}) or (\ref{frolovmatrix}). Also note that there is
no B-field in the original background so the RR fields $D_m$
defined in (\ref{RR1}) are simply equal to $C_m$. The operator
corresponding to our solution generating matrix, which acts on the
Fock space, was constructed in \cite{oota} as
\begin{equation}\label{RR6}
{\bf \Lambda} = \exp(\frac{1}{2} \Gamma_{mn} \psi_m \psi_n).
\end{equation}
From the discussion given above one can see easily that the
corresponding operator acting on the differential forms is
\begin{equation}\label{RR7}
{\bf T} = \exp(\frac{1}{2} \Gamma_{mn} i_m i_n),
\end{equation}
where $i_m$ is contraction with  the isometry direction
$\partial/\partial y^m$.

In all the $AdS_5 \times X_5$ backgrounds that are of interest to
us here, the only RR field is $C_4$ with field strength
\begin{equation}\label{RR8}
F_5 = \frac{(16 \pi N)\pi^3}{{\rm Vol}(X_5)}(w_{AdS_5} +
*w_{AdS_5})
\end{equation}
where $w_{AdS_5}$ is the volume form on $AdS_5$ and $*$ is the
Hodge dual, so that $*w_{AdS_5}$ is the volume form on $X_5$. For
example when $X_5 = Y^{p,q}$ we have
$$*w_{AdS_5} = \frac{l}{18} (1-y)s_{\theta}d\theta \wedge dy \wedge
d\psi \wedge d\alpha \wedge d\phi.$$ So, the RR fields of the new
background obtained by the solution generating matrix
(\ref{matrix}) is
\begin{equation}\label{RR9}
\frac{16\pi N}{V}(1-\gamma i_{\alpha} i_{\phi})(w_{AdS_5} +
*w_{AdS_5}) = \frac{16\pi N}{V}((w_{AdS_5} + *w_{AdS_5})+
\frac{\gamma l}{18}(1-y)s_{\theta} d\theta \wedge dy \wedge
d\psi).
\end{equation}
Here we have used
$$ V^{-1} = \frac{{\rm Vol}(S^5)}{{\rm Vol}(Y^{p,q})} =
\frac{\pi^3}{{\rm Vol}(Y^{p,q})} = \left[\frac{q^2[2p + (4p^2 -
3q^2)^{1/2}]}{3p^2[3q^2 - 2p^2 + p(4p^2 -
3q^2)^{1/2}]}\right]^{-1}.$$ In a suitable gauge we obtain
\begin{eqnarray}
F_5 &=& \frac{16 \pi N}{V}(w_{AdS_5} + *w_{AdS_5}) \nonumber \\
C_2 &=&-\frac{8 \pi N \gamma l}{9 V}(1-y)c_{\theta}dy \wedge
d\psi. \nonumber
\end{eqnarray}
This is the RR fields of the one-parameter deformed background of
\cite{LM}\footnote{Note that there is a sign flip for $C_2$. As
has already been noted in \cite{F}  the sign conventions of
\cite{LM} are different from the literature.}. The generalization
to our case is now obvious and we obtain
\begin{eqnarray}\label{RR10}
F_5 &=& \frac{16 \pi N}{V}(w_{AdS_5} + *w_{AdS_5}) \nonumber \\
C_2 &=&-\frac{8 \pi N  l}{9 V}(1-y)c_{\theta}(\gamma_3  dy \wedge
d\psi + \gamma_2  dy \wedge d\phi + \gamma_1  dy \wedge d\alpha).
\end{eqnarray}

\noindent On the other hand, for the $T^{1,1}$ case the volume
form and the total volume are $$ *w_{AdS_5} =
\frac{1}{108}s_{\theta_1} s_{\theta_2} d\theta_1 \wedge d\theta_2
\wedge d\phi_1  \wedge d\phi_2 \wedge d\psi \ \ \ {\rm and} \ \ \
{\rm Vol}(T^{1,1}) = \frac{16 \pi^3}{27}.$$ The results of
\cite{LM} is obtained from
\begin{equation}\label{RR11}
27 \pi N(1-\gamma i_{\phi_1} i_{\phi_2})(w_{AdS_5} + *w_{AdS_5}) =
27 \pi N((w_{AdS_5} + *w_{AdS_5})+ \gamma \frac{s_1 s_2}{108}
d\theta_1 \wedge d\theta_2 \wedge d\psi),
\end{equation}
so that in a suitable gauge one has
\begin{equation}\label{RR12}
F_5 =27 \pi N(w_{AdS_5} + *w_{AdS_5}), \ \ \ \ \ C_2 =
-\frac{\gamma \pi N}{4} c_1 s_2 d\theta_2 \wedge d\psi.
\end{equation}
The generalizaton to the 3-parameter case is now obvious and here
we  present the result as
\begin{eqnarray}\label{RR15}
F_5 &=& 27 \pi N(w_{AdS_5} + *w_{AdS_5}) \nonumber \\
C_2 &=& -\frac{\pi N}{4} c_1 s_2 (\gamma_3 d\theta_2 \wedge d\psi
+ \gamma_2 d\theta_2 \wedge d\phi_2 + \gamma_1 d\theta_2 \wedge
d\phi_1).
\end{eqnarray}


\section{Regularity of Solutions}

In this section, we analyze the regularity of the new solutions.
As was already mentioned in the text, under a T-duality
transformation of the form (\ref{onemli}), the metric always
transforms as $$g \longrightarrow g' = \frac{1}{(CE+D)^t} \ g \
\frac{1}{(CE+D)}$$ so that the transformation of the volume form
is
\begin{equation}\label{volume}
* {\bf 1} = \sqrt{{\rm det}g} \ dy_1 \wedge dy_2 \wedge dy_3
\longrightarrow \sqrt{{\rm det}g'} \ dy_1 \wedge dy_2 \wedge dy_3
=G \sqrt{{\rm det}g} \ dy_1 \wedge dy_2 \wedge dy_3 = *' {\bf 1}
\end{equation}
Here $y_i$ are the coordinates of the 3-torus and $*'$ is the
Hodge star operator of the deformed geometry. The form of $G =
{\rm det}(CE+D)^{-1}$ for a general 3-parameter deformation, for
which $D=I$ and $C = \Gamma$ in (\ref{frolovmatrix}), was given in
(\ref{G}). First thing we observe is that $G$ has no zeroes, so
starting with a non-singular solution, we end up with a
non-singular solution, unless $G$ blows up at some points. In
general, as $\Delta_i$ and $K_i$ can become negative, $G$ may have
poles for some backgrounds, when $\gamma_i$ are related. Let us
make a case by case analysis for the examples considered here. In
the case of $AdS_5 \times S^5$, $K_i = 0$ and $\Delta_i \geq 0$,
so $G$ does not have any poles in either of the one-parameter and
three-parameter deformations. For one-parameter deformations of
$T^{1,1}$ and $Y^{p,q}$, for which $\gamma_1 = \gamma_2 = 0$,
there is no contribution from $K_i$, and $\Delta_3 \geq 0$, so
again, $G$ has no poles. Now consider the new three-parameter
deformation of $T^{1,1}$, which  we gave in (\ref{newt11}). Here
too, $G$ has no poles, as can be seen easily when written in the
following form:
$$G^{-1} = \frac{1}{108}(108 +
2s_{\theta_1}^2(c_{\theta_2} \gamma_3 - \gamma_2)^2 +
2s_{\theta_2}^2(c_{\theta_1} \gamma_3 - \gamma_1)^2 + 3 \gamma_3^2
s_{\theta_1}^2 s_{\theta_2}^2).$$ For the analysis of $Y^{p,q}$,
one should first notice that the range of $y$ is fixed such that
$1-y > 0, \ a-y^2 > 0, \ w(y) > 0, \ q(y) \geq 0$ \cite{MSJ}. From
this, one immediately sees that $\Delta_i \geq 0$. However, $K_1$
and $K_2$ can become negative at some points. For example, $K_1<0$
when $\pi/2 < \theta < \pi$. So, for some points or on some lines
or planes in the parameter space, $G$ might have poles at some
values of $\theta$ and $y$. Note that the plane $\gamma_3=0$ is
safe in the parameter space, as then there is no contribution from
$K_1$ or $K_2$.

The next thing to be noticed about the new three-parameter
deformations found here is that it does not reduce to the
undeformed solution when the volume form of the three-torus goes
to zero. As can be seen from (\ref{100}), when $e=0$, the metric
scales as $g_{ij} \rightarrow G g_{ij}$ and the B-field does not
vanish. However, when the volume forms of all the three 2-tori
within the 3-torus are zero, that is, when $\Delta_1 = \Delta_2 =
\Delta_3 = 0$ (which immediately implies that $K_1=K_2=K_3=e=0$),
our solution reduces to the undeformed solution.

Finally, an important criteria for the regularity of the new
solutions is that the RR field $C_2$ should vanish, as the
internal torus shrinks to zero size \cite{LM}. In the previous
section, we have seen that $F_3=dC_2$ is schematically of the form
\begin{equation}\label{C2}
F_3 = \frac{(16 \pi N)\pi^3}{{\rm Vol}(X_5)} \sqrt{g(X_5)} \ dy_1
\wedge dy_2 \wedge \sum_{i=1,2,3}\gamma_i dy_i
\end{equation}
where $\sqrt{g(X_5)} \ dy_1 \wedge dy_2 \wedge dy_3 \wedge dy_4
\wedge dy_5$ is the volume form of $X_5$ and $dy_i, i=1,2,3$ are
the coordinates of the 3-torus of isometries, $X_3$. So the
behavior of $C_2$ as the 3-torus shrinks depends entirely on how
the volume form of $X_3$ is related to that of $X_5$. In the case
of $T^{1,1}$, $\sqrt{g(X_5)} = \frac{1}{108}
s_{\theta_1}s_{\theta_2} = \frac{1}{36} \sqrt{g(X_3)}$, so $F_3$
vanishes  at the zeroes of the volume form of $X_3$. At these
points $C_2$ is pure gauge and as such it can be gauged away.
However, in the case of $Y^{p,q}$
$$\sqrt{g(X_5)} = \frac{l(1-y)s_{\theta}}{18}, \ \ \ \ \
\sqrt{g(X_3)}= \frac{l s_{\theta}}{3} \sqrt{\frac{2y^3 - 3y^2 +
a}{3}}.$$ We see that $C_2$ does not necessarily vanish as the
3-torus of isometries shrinks to zero size.


\section{Conclusions}

In this paper, we have examined the solution generating symmetries
used by \cite{LM} and identified the $O(2,2,\IR)$ matrix that acts
on $E=g+B$. The solution generating $O(2,2,\IR)$ acts locally on
the moduli space of two dimensional conformal field theories on
the world-sheet, taking one CFT to another connected to it by an
exactly marginal deformation. Correspondingly, the dual field
theory on the boundary of $AdS_5$ is deformed by an exactly
marginal operator, giving rise to  $\beta$ deformations of the
original theory. On the other hand, the discrete $O(2,2,Z)$ takes
the world-sheet CFT to an equivalent one, just like the
corresponding deformation of the field theory with  integer
$\beta$ gives the same theory that one has started with. We have
seen that some of the important features of the deformed gravity
backgrounds follow from the properties of the $O(2,2,\IR)$ matrix.

We  found here a new 3-parameter family of deformations of the
Sasaki-Einstein manifolds $T^{1,1}$ and $Y^{p,q}$. We have seen
that the new solutions associated with $T^{1,1}$ are regular,
whereas there are some problems with the regularity of the
deformations of $Y^{p,q}$.

An obvious next step would be to find the Lax pair of these new
backgrounds following the methods developed by \cite{F}. It would
also be interesting to examine  the associated spin chain as was
done in \cite{beisert} for the $AdS_5 \times S^5$ case.


\section*{Acknowledgements}
I am grateful to Sergey Cherkis for  helpful discussions, for
reading  the manuscript and for useful  suggestions. I also would
like to thank Chris Hull and Dan Waldram for helpful e-mail
correspondence.  This work is supported by Irish Research Council
for Science, Engineering and Technology (IRCSET) under the
postdoctoral fellowship scheme.

\typeout{LaTeX Warning: Label(s) may have changed. Rerun}

\end{document}